\pdfoutput=1
\documentclass[a4paper,11pt]{article}
\usepackage{pos}
\usepackage{float}
\usepackage{mathtools}
\usepackage{iftex}
\usepackage{tikz}
\usetikzlibrary{positioning}
\iftutex
  \usepackage{fontspec}
\fi

\title{The DIS structure functions $F_4$ and $F_5$ at NLO and their impact on $\tau$-neutrino scattering}
\ShortTitle{$F_4$ and $F_5$ at NLO and $\tau$-neutrino scattering}

\author*[a]{T.~Je\v{z}o}
\author[a]{M.~Klasen}
\author[b,c]{P.~Risse}
\author[d]{I.~Schienbein}
\author[a]{E.~Spezzano}

\affiliation[a]{Institut  für  Theoretische  Physik,  Universität  Münster,  Wilhelm-Klemm-Straße 9, 48149 Münster, Germany}
\affiliation[b]{Department of Physics, Southern Methodist University, Dallas, TX 75275-0175, U.S.A.}
\affiliation[c]{Theory Center, Jefferson Lab, Newport News, VA 23606, U.S.A.}
\affiliation[d]{Laboratoire de Physique Subatomique et de Cosmologie, Université Grenoble-Alpes,CNRS/IN2P3, 53 Avenue des Martyrs, 38026 Grenoble, France}

\emailAdd{tomas.jezo@uni-muenster.de}

\abstract{
We present next-to-leading-order calculations of heavy-quark contributions to the deep-inelastic scattering structure functions $F_4$ and $F_5$ within the Aivazis--Collins--Olness--Tung scheme, implemented in the open-source library \texttt{APFEL{++}} and evaluated using \texttt{CT18NLO} parton distribution functions \cite{Spezzano:2026jhep}. 
These structure functions are suppressed by lepton mass effects in light-lepton processes, but become important in $\tau$-lepton and neutrino scattering at facilities such as SHiP, IceCube, and DUNE. We furthermore quantify the direct impact of $F_4$ and $F_5$ on the charged-current $\tau$-neutrino cross section, for the first time presented here, showing they can suppress it by up to $-40\%$ near the kinematic boundary at low $Q^2$ and large $x$.
}

\renewcommand{\hookAfterAbstract}{%
\begin{tikzpicture}[remember picture,overlay]
    \node [below left=5mm and 4mm of current page.north east] (preprint) {MS-TP-26-27};
\end{tikzpicture}%
}

\FullConference{33rd International Workshop on Deep Inelastic Scattering and Related Subjects (DIS2026)\\
4--8 May 2026\\
Bologna, Italy\\}

\begin{document}
\maketitle

\section{Introduction}
The structure functions $F_1$, $F_2$, and $F_3$ of deep-inelastic scattering (DIS) are well-studied both theoretically and experimentally, while $F_4$ and $F_5$ have never been measured due to suppression by small size of the lepton mass \citep{ParticleDataGroup:2024cfk}. 
They become important in $\tau$-lepton and $\tau$-neutrino scattering accessible at SHiP \citep{shipcollaboration2022shipexperimentproposedcern}, IceCube \citep{IceCube:2024nhk}, and DUNE \citep{abi2020deepundergroundneutrinoexperiment}. Contributions due to heavy quarks play a critical role, especially at low Bjorken-$x$, where gluon and sea-quark distributions dominate \citep{Stavreva:2012bs}. The Aivazis--Collins--Olness--Tung (ACOT) scheme \citep{Aivazis:1993kh,Aivazis:1993pi} consistently incorporates heavy-quark mass effects within a variable-flavor-number framework, extending prior calculations for $F_1$, $F_2$, $F_3$ \citep{Kretzer:1998ju} to $F_4$ and $F_5$. We have implemented these calculations at next-to-leading (NLO) QCD accuracy in \texttt{APFEL{++}} \citep{Bertone:2016lga} end numerically evaluated them using \texttt{CT18NLO} PDFs \citep{Hou_2021}; full details are in the companion paper \citep{Spezzano:2026jhep}.

In this contribution, we briefly summarize the theoretical framework and present representative numerical results for the $F_4$ and $F_5$ structure functions.
In addition, we present new numerical results for the impact of $F_4$ and $F_5$ on charged-current $\tau$-neutrino DIS cross sections, quantifying the regions of phase space in which these otherwise suppressed structure functions become phenomenologically relevant.

\section{Cross sections and structure functions}
The charged-current (CC) differential cross section for $\tau$-neutrino DIS reads \citep{ParticleDataGroup:2024cfk}:
\begin{align}
\frac{d^2\sigma^{W}}{dx\,dy} &= \frac{4\pi\alpha^2}{x y Q^2} \tilde{\eta}\eta^W \Bigg\{ \,
\frac{x y^2 \big(Q^2 + \mu_1^2 + \mu_2^2\big)}{Q^2} \, F_1^{W}  - \frac{2 \Big( (y-1) Q^4 + M^2 x^2 y^2 (Q^2 + \mu_1^2 + \mu_2^2) \Big)}{Q^4} \, F_2^{W} \nonumber \\[4pt]
&\quad \mp\frac{x y \big( (2-y)Q^2 + y (\mu_1^2- \mu_2^2) \big)}{Q^2} \, F_3^{W}  + \frac{x y^2 \Big( \mu_1^4 + (Q^2 - 2 \mu_2^2)\mu_1^2 + \mu_2^2(Q^2 + \mu_2^2) \Big)}{Q^4} \, F_4^{W} \nonumber \\[4pt]
&\quad - \frac{2x y \big( (y-1)\mu_1^2 + \mu_2^2 \big)}{Q^2} \, F_5^{W}
\Bigg\},
\label{eqn: CC cross section}
\end{align}
where $F_i$ are structure functions, $Q$, $x$ and $y$ are the usual kinematic variables used in DIS. $\mu_1$ and $\mu_2$ are the incoming and outgoing lepton masses, $\tilde{\eta}=2$ for incoming (anti)neutrinos, and $\eta^W$ is the $W$-boson propagator. The $F_4$ and $F_5$ contributions are suppressed by powers of $\mu_2^2/Q^2$, making them measurable only in processes involving a lepton with non-negligible mass such as the $\tau$. An analogous expression for neutral-current differential cross section can be found e.g.~in Ref.~\cite{Spezzano:2026jhep}. 

The structure functions are calculated in perturbative QCD at NLO using the ACOT scheme. 
This includes the Born-level quark-scattering contribution, its real and virtual $\mathcal{O}(\alpha_s)$ corrections, and the boson--gluon-fusion channel, retaining the full dependence on the relevant quark masses in the hard-scattering coefficient functions. 
To avoid double counting, the quasi-collinear logarithms already contained in the DGLAP-evolved heavy-quark PDFs are removed through the corresponding ACOT subtraction terms. 
The resulting description therefore smoothly interpolates between the massive fixed-flavour treatment near the heavy-quark threshold and the massless variable-flavour result at $Q^2 \gg m_Q^2$, where the logarithms $\ln(Q^2/m_Q^2)$ are resummed through PDF evolution.

\section{Numerical results}
We now turn to the numerical results. 
We first examine the behaviour of the newly calculated structure functions $F_4$ and $F_5$, illustrating their dependence on $x$ and $Q^2$ and the size of the NLO corrections. 
We then go beyond the structure-function level and study their impact on charged-current $\tau$-neutrino DIS, quantifying the resulting modification of the differential cross section across the accessible phase space.

\begin{figure}[H]
    \centering
    \includegraphics[width=0.45\textwidth]{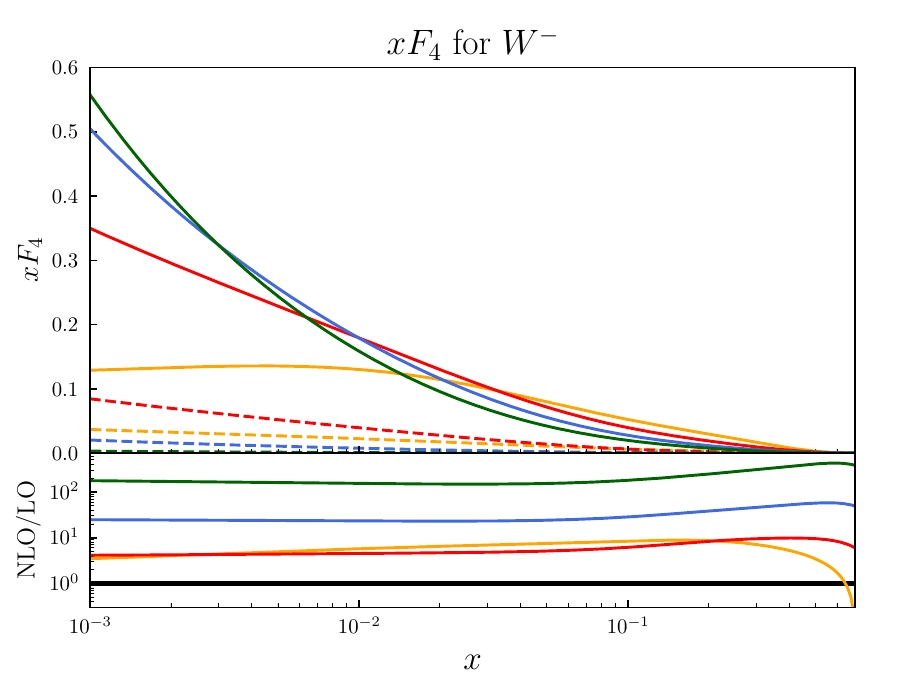}%
    \includegraphics[width=0.45\textwidth]{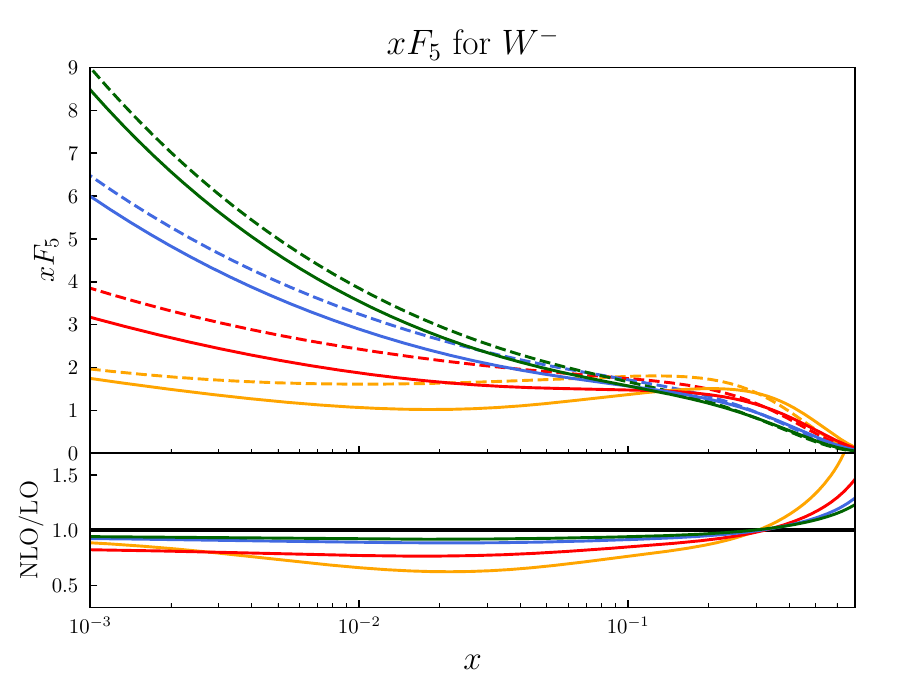}
    \caption{The CC structure functions $xF_4$ (left) and $xF_5$ (right) for $W^-$ exchange, computed in the ACOT scheme using CT18NLO PDFs. Absolute predictions at LO (dashed) and NLO (solid) are shown for $Q^2 \in \{2,\,10,\,100,\,1000\}\,\mathrm{GeV}^2$ (yellow, red, blue, green). The lower panels show the NLO/LO ratios (K-factors) across $x \in [10^{-3}, 0.7]$.}
    \label{fig:F4F5}
\end{figure}

As shown in Fig.~\ref{fig:F4F5}, $F_4$ and $F_5$ exhibit rather different perturbative behaviour.
The LO contribution to $F_4$ is strongly suppressed and vanishes in the massless-quark limit. 
More generally, the Albright--Jarlskog (AJ) relation gives $2xF_4=F_L$ in this limit, so that $F_4$ is naturally generated at $\mathcal{O}(\alpha_s)$, where it is dominated by the gluon-initiated contribution. 
The large NLO/LO ratios observed for $F_4$ should therefore not be interpreted as an unusually large conventional $K$-factor, but rather reflect the mass suppression of the Born contribution.

In contrast, $F_5$ receives a sizeable quark-initiated contribution already at LO, with a characteristic valence-quark enhancement at large $x$. 
For $W^-$ exchange this feature is particularly pronounced due to the dominant up-quark contribution. 
The NLO corrections are considerably more moderate than for $F_4$, although deviations from the LO result can still reach ${\sim}45\%$. 
At small $x$, the increasing importance of sea quarks and gluons reduces the sensitivity to the valence-flavour structure. 
In the large-$Q^2$ limit, heavy-quark mass effects diminish and the results approach the
AJ relations, $2xF_4=F_L$ and $xF_5=F_2$. 
Results for $W^+$ exchange and the neutral-current channel are presented in Ref.~\citep{Spezzano:2026jhep}.

To quantify the phenomenological relevance, we define the ratio of the full cross section to the one with $F_4=F_5=0$:
\begin{equation}
 R(x, y, Q^2) = \frac{\frac{d^2\sigma^{W^+}}{dx \, dy}}{\frac{d^2\sigma^{W^+}}{dx \, dy}\mathrlap{\left. \right|_{F_4=F_5=0}}} =1+\delta R(x, y, Q^2)
  \label{eq:R_ratio}
\end{equation}
where $\delta R \equiv \mathcal{A}/\mathcal{B}$, with $\mathcal{A}$ the $F_4, F_5$ terms and $\mathcal{B}$ the baseline $F_1, F_2, F_3$ terms of Eq.~\eqref{eqn: CC cross section} (setting $\mu_1=0$, $\mu_2=m_\tau$). The kinematically allowed region satisfies $0\le y\le Q^2/(Q^2+m_\tau^2)$ \citep{Kretzer_2002}.

\begin{figure}[H]
  \centering
  \includegraphics[width=0.9\textwidth]{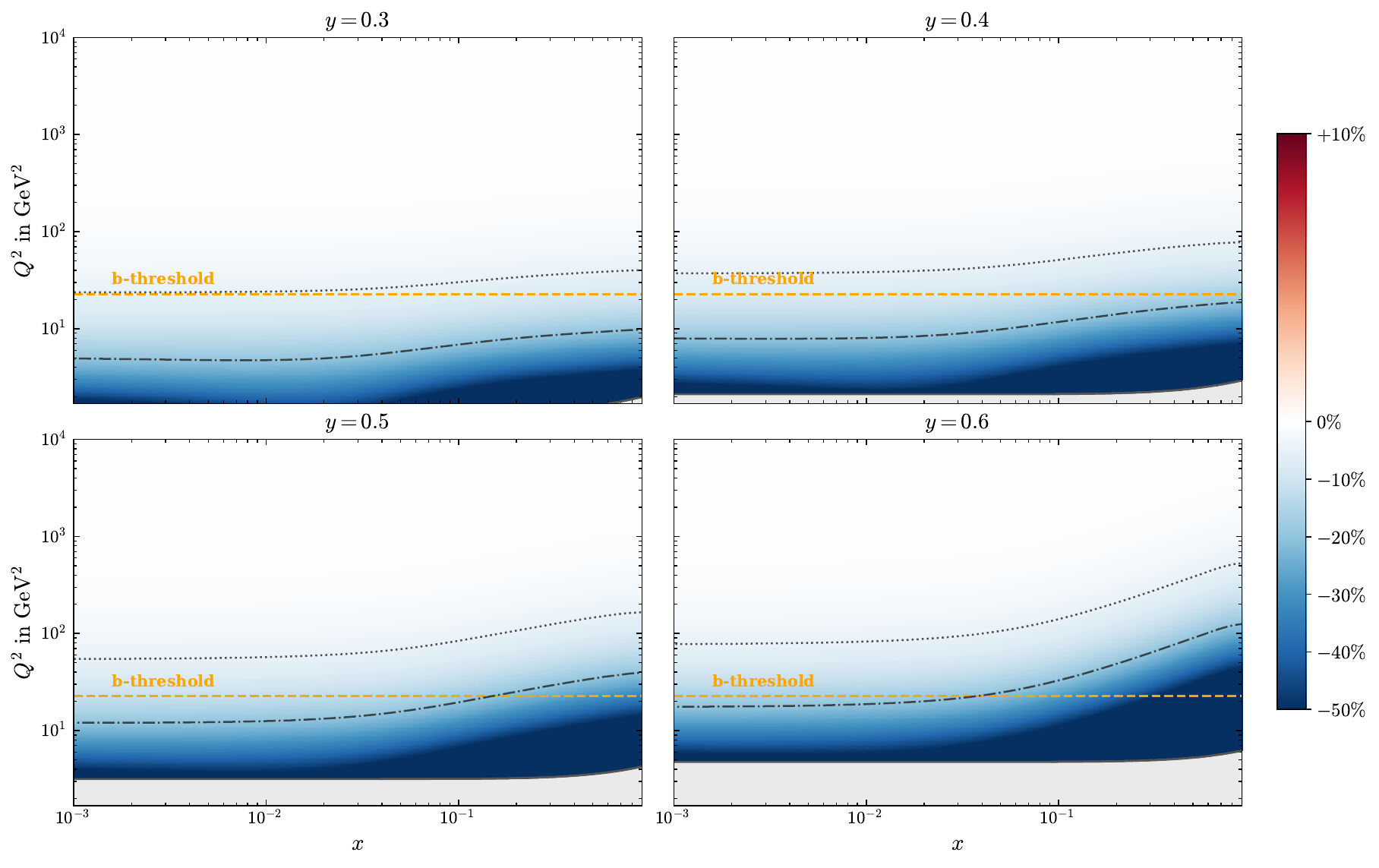}
  \caption{Relative cross-section change $\delta R$ due to $F_4$ and $F_5$ in the $W^+$ CC $\tau$-neutrino cross section, for $y=0.3,\,0.4,\,0.5,\,0.6$. The grey region is kinematically forbidden; dash-dotted and dotted contours mark $-20\%$ and $-5\%$ levels. The dashed orange line indicates $Q^2=m_b^2$.}
  \label{fig:cc_ratio}
\end{figure}

The ratio $\delta R$ is negative throughout the physical domain, confirming a suppression from $F_4$ and $F_5$. The suppression reaches up to $-40\%$ near the kinematic boundary at low $Q^2$ and large $x$ and gets more pronounced with $y$. 
The $-20\%$ contour lies at $Q^2\lesssim 10$--$30\,\mathrm{GeV}^2$ for $x\gtrsim 0.1$--$0.3$, and the $-5\%$ contour extends to $Q^2\approx 80\,\mathrm{GeV}^2$ for $y=0.6$. 
For $Q^2\gtrsim 10^3\,\mathrm{GeV}^2$ corrections fall below $1\%$, as expected from heavy lepton suppression.

\section{Summary}

We have computed NLO heavy-quark contributions to $F_4$ and $F_5$ in the ACOT scheme \citep{Spezzano:2026jhep}, implemented in \texttt{APFEL{++}}. 
$F_4$ is dominated by NLO corrections (vanishing LO in the massless limit via the AJ relations), while $F_5$ receives moderate NLO corrections up to $\sim 45\%$ in the CC case. 
Going beyond the structure-function results of Ref.~\citep{Spezzano:2026jhep}, we have quantified the impact of $F_4$ and $F_5$ on charged-current $\tau$-neutrino DIS. 
The combined contribution of $F_4$ and $F_5$ suppress the CC $\tau$-neutrino cross section throughout the physical phase space, with
corrections reaching up to $-40\%$ near the kinematic boundary low $Q^2$ and large $x$.
These results demonstrate that the $F_4$ and $F_5$ must be included consistently in precision predictions for $\tau$-neutrino DIS in the kinematic regions relevant to experiments such as DUNE, SHiP, and
IceCube.

\section*{Acknowledgments}

The speaker thanks the organizers for the kind invitation and for providing a platform for a fruitful exchange of ideas, and his nCTEQ colleagues for insightful discussions.
This work has been supported by the BMBF under contract 05P24PMA.
The work of P.R. was supported by the U.S. Department of Energy under Grant No. DE-SC0010129, and by the Office of Science, the Office of Nuclear Physics, within the framework of the Saturated Glue (SURGE) Topical Theory Collaboration. This material is based upon work supported by the U.S. Department of Energy, Office of Science, Office of Nuclear Physics under contract No. 89243126CS000213.

\bibliographystyle{JHEP}
\bibliography{polACOT_references}

\providecommand{\href}[2]{#2}\begingroup\raggedright\begin{thebibliography}{10}

\bibitem{Spezzano:2026jhep}
E.~Spezzano, T.~Je{\v z}o, M.~Klasen, P.~Risse, and I.~Schienbein, {\it
  {Heavy-quark contributions to the DIS structure functions $F_{4}$ and $F_{5}$
  at NLO in the ACOT scheme}},  {\em JHEP} {\bf 05} (2026) 012,
  [\href{http://arxiv.org/abs/2512.11569}{{\tt arXiv:2512.11569}}].

\bibitem{ParticleDataGroup:2024cfk}
{\bf Particle Data Group} Collaboration, S.~Navas et~al., {\it {Review of
  particle physics}},  {\em Phys. Rev. D} {\bf 110} (2024), no.~3 030001.

\bibitem{shipcollaboration2022shipexperimentproposedcern}
{SHIP Collaboration}, {\it The ship experiment at the proposed cern sps beam
  dump facility},  2022.

\bibitem{IceCube:2024nhk}
{\bf IceCube} Collaboration, R.~Abbasi et~al., {\it {Observation of Seven
  Astrophysical Tau Neutrino Candidates with IceCube}},  {\em Phys. Rev. Lett.}
  {\bf 132} (2024), no.~15 151001, [\href{http://arxiv.org/abs/2403.02516}{{\tt
  arXiv:2403.02516}}].

\bibitem{abi2020deepundergroundneutrinoexperiment}
{B. Abi et al.}, {\it {Deep Underground Neutrino Experiment (DUNE), Far
  Detector Technical Design Report, Volume II: DUNE Physics}},  2020.

\bibitem{Stavreva:2012bs}
T.~Stavreva, F.~I. Olness, I.~Schienbein, T.~Jezo, A.~Kusina, K.~Kovarik, and
  J.~Y. Yu, {\it {Heavy Quark Production in the ACOT Scheme at NNLO and N3LO}},
   {\em Phys. Rev. D} {\bf 85} (2012) 114014,
  [\href{http://arxiv.org/abs/1203.0282}{{\tt arXiv:1203.0282}}].

\bibitem{Aivazis:1993kh}
M.~A.~G. Aivazis, F.~I. Olness, and W.-K. Tung, {\it {Leptoproduction of heavy
  quarks. 1. General formalism and kinematics of charged current and neutral
  current production processes}},  {\em Phys. Rev. D} {\bf 50} (1994)
  3085--3101, [\href{http://arxiv.org/abs/hep-ph/9312318}{{\tt
  hep-ph/9312318}}].

\bibitem{Aivazis:1993pi}
M.~A.~G. Aivazis, J.~C. Collins, F.~I. Olness, and W.-K. Tung, {\it
  {Leptoproduction of heavy quarks. 2. A Unified QCD formulation of charged and
  neutral current processes from fixed target to collider energies}},  {\em
  Phys. Rev. D} {\bf 50} (1994) 3102--3118,
  [\href{http://arxiv.org/abs/hep-ph/9312319}{{\tt hep-ph/9312319}}].

\bibitem{Kretzer:1998ju}
S.~Kretzer and I.~Schienbein, {\it {Heavy quark initiated contributions to deep
  inelastic structure functions}},  {\em Phys. Rev. D} {\bf 58} (1998) 094035,
  [\href{http://arxiv.org/abs/hep-ph/9805233}{{\tt hep-ph/9805233}}].

\bibitem{Bertone:2016lga}
V.~Bertone, S.~Carrazza, and N.~P. Hartland, {\it {APFELgrid: a high
  performance tool for parton density determinations}},  {\em Comput. Phys.
  Commun.} {\bf 212} (2017) 205--209,
  [\href{http://arxiv.org/abs/1605.02070}{{\tt arXiv:1605.02070}}].

\bibitem{Hou_2021}
T.-J. Hou, J.~Gao, T.~Hobbs, K.~Xie, S.~Dulat, M.~Guzzi, J.~Huston,
  P.~Nadolsky, J.~Pumplin, C.~Schmidt, I.~Sitiwaldi, D.~Stump, and C.-P. Yuan,
  {\it New cteq global analysis of quantum chromodynamics with high-precision
  data from the lhc},  {\em Physical Review D} {\bf 103} (Jan., 2021).

\bibitem{Kretzer_2002}
S.~Kretzer and M.~H. Reno, {\it Tau neutrino deep inelastic charged current
  interactions},  {\em Physical Review D} {\bf 66} (Dec., 2002).

\end{thebibliography}\endgroup

\end{document}